\documentstyle[12pt,aaspp4,psfig]{article}

\begin{document}

\title{The Effects of Low-Temperature Dielectronic Recombination on the Relative Populations of the Fe M-Shell States}

\author{S. B. Kraemer\altaffilmark{1},
G.J. Ferland\altaffilmark{2},
\& J.R. Gabel\altaffilmark{1}}

\altaffiltext{1}{Catholic University of America,
and Laboratory for Astronomy and Solar Physics, NASA's Goddard Space Flight 
Center, Code 681,
Greenbelt, MD  20771; kraemer@yancey.gsfc.nasa.gov, gabel@iacs.gsfc.nasa.gov}

\altaffiltext{2}{Dept. of Physics and Astronomy, University of Kentucky,
Lexington, KY, 40506; gary@pa.uky.edu}

\begin{abstract}

We examine the effects of low-temperature, or $\Delta$$n = 0$, dielectronic recombination (DR) on
the ionization balance of the Fe M-Shell (Fe~IX through Fe~XVI). Since
$\Delta$$n = 0$ rates are not 
available for these ions, we have derived estimates based on the existing rates for the first four ionization
states of the CNO sequence and newly calculated rates for L-shell ions of 3rd row elements and Fe.
For a range of ionization parameter and column density 
applicable to the intrinsic absorbers detected in {\it ASCA}, {\it Chandra}, and {\it XMM-Newton} observations of 
Seyfert galaxies, we generated two grids of photoionization models, with and without DR. The results show that the ionization 
parameter at which the population of an Fe M-shell ion peaks
can increase in some cases by factor $>$ 2 when these rates are included. More importantly, there are dramatic changes in the
range in ionization parameter over which individual M-shell ions contain significant fractions of the total Fe (e.g. $>$ 10\%) in the plasma.
These results may explain
the mismatch between the range of Fe ionization states detected in the X-ray spectra of Seyferts, identified by the energies of the
M-Shell Unresolved Transition Array, and those predicted by photoionization models of the X-ray absorbers that reproduce
lines of second and third row elements. The results
suggest that care should be taken in using 3rd and 4th row ions to constrain the physical conditions in photoionized X-ray plasmas
until accurate DR rates are available. This underscores the importance of atomic physics in interpreting
astronomical spectroscopy.
       
\end{abstract}


\section{Introduction}

New X-ray observatories such as the {\it Chandra X-ray Observatory} and {\it XMM- Newton} have made high
resolution spectroscopy of active galaxies possible for the first time, allowing the detection of phases of gas that were previously
unknown. The physical state of this gas is interpreted by reference to large-scale numerical
simulations, hence, it is critical that physical processes that occur in an ionized plasma be fully understood
(the main topic of the 
conference ``Spectroscopic Challenges of
Photoionized Plasmas'', held at the University of Kentucky, Lexington, Kentucky, USA on the 15 - 18 November, 2000;  
[Ferland
\& Savin 2001]).

 The ionization structure and temperatures in cosmic plasmas depend strongly on ionization and recombination processes. Recombination
can occur via radiative recombination, charge transfer, or dielectronic
recombination (DR). The DR process involves the
capture of a free electron by an ion, resulting in doubly-excited recombined species with the electron in an auto-ionizing state.
This is followed either by auto-ionization (i.e., the reverse process) or the emission of a photon, the latter leaving the recombined ion
in a stable state. 

There are two varieties of DR. The first involves the capture of the electron to auto-ionizing states
with high principal quantum number, with an accompanying core excitation with a change in principal quantum number
$\Delta$$n > 0$ (e.g. Burgess 1964). This form of DR dominates over radiative recombination at temperatures comparable
with the ionization potential, i.e collisionally ionized plasmas.
Nevertheless, terms of the same parity as the ground term exist between
the ground state and the first excited state of the opposite parity, and thus a series of autoionizing states converge on these terms
(see Nussbaumer \& Storey 1983). This results in a second type of DR in which there is no change in the principal quantum number of the excited core,
 or $\Delta$$n = 0$,  
which can proceed at low temperatures for which the thermal energy of the free electron is less than the excitation energy of 
first resonance of the recombining ion. Low-temperature DR is usually the dominant process in  
photoionized plasmas, with temperatures typically 5 - 10\% of the ionization potential, such as those detected in X-ray spectra of Seyfert galaxies 
(e.g Reynolds 1997; George et al. 1998; Sako et al. 2000; Kaspi et al. 2001, 2002; Kinkhabwala et al. 2002; Turner et al. 2003). 

While low-temperature DR
rates have been calculated for a number of abundant 2nd and a few 3rd row elements (Nussbaumer \& Storey 1983; Nussbaumer \& Storey 1986; Nussbaumer \& Storey 1987),
the full set of ionization states have only been completed for 2nd row elements. One of the difficulties in these calculations is that the 
rates depend on the accurate determination of the low-level autoionizing states (see Gu 2003). Ali et al. (1991) estimated
low-temperature DR rate coefficients for other ions by taking the mean values from the first four ions of
C, and O calculated by Nussbaumer \& Storey (1983), and demonstrated that the model-predicted emission-line
strengths were strongly affected when these rates were introduced. Savin (2000) has shown that uncertainties in DR rates 
translate into similar uncertainties in predicted ionic column densities, which may hamper the determination of elemental abundances ratios
in the intergalactic medium.
There has been recent work in determining both high- and low-
temperature DR rates for L-shell (the Li-
through Ne-like iso-electronic sequences), both experimentally (Savin et al. 1999, 2002) and computationally (Gu 2003;
Badnell et al. 2003), the results of which emphasize the importance of this process in photoionized plasma. 
Note that, due to the difficulties in calculating the energies of the autoionizing states, experimental results are
critical.
 
On the other hand, there has not been any recent work published on $\Delta$$n = 0$ rates for M-shell ions (the Na- to Ar-like sequence).
This is unfortuitous, since M-shell Fe ions, Fe~IX to Fe~XVI, contribute strongly to the unresolved transition array (UTA) of inner
shell $n = 2-3$ lines seen in absorption in high resolution X-ray spectra of several Seyfert galaxies observed with {\it XMM-Newton} and {\it Chandra}
such as IRAS 13349+2438 (Sako et al. 2001), NGC 3783 (Blustin et al. 2002; Kaspi et al. 2002; Netzer et al. 2003), Mrk 509
(Pounds et al. 2001; Yaqoob et al. 2003), and NGC 5548 (Steenbrugge et al. 2003; Steenbrugge et al., in preparation). For N- and M-shell
Fe, the UTA features lie between $\sim$ 15.5 \AA~ and 17 \AA~, depending on which ions are present (Behar, Sako, \& Kahn 2001). In principal, the UTA
can be used to constrain the ionization state of the absorbers, which is especially valuable if the resonance lines from H-, He-,
and Li-like C, N, and O are saturated. However, in the case of NGC 3783  (Netzer et al. 2003), the ionization state suggested by the UTA
wavelengths seems lower than that predicted by photoionization models generated to fit CNO absorption features and
the ionization state/column density of 3rd row elements such as Si and S. A similar discrepancy appears to be the case for
NGC 5548 (Steenbrugge et al., in preparation). One of the possible causes Netzer et al. suggested to explain this mismatch is the
lack of $\Delta$$n = 0$ rates for
the M-shell Fe ions, which leads to the models predicting the Fe in too high an ionization state (compared to lower Z elements).      

In order to explore the effect of $\Delta$$n = 0$ DR on the M-shell Fe ionization balance, we adopted the approach described in
Ali et al. (1991), adding estimated rate coefficients to the development version of the photoionization code Cloudy (Ferland et al. 1998). In order  to
gauge the magnitude of the effect, model grids were generated with and without these rates. We also examined the effects of
uncertainties in the estimates rates. The model details and results
are described below. 

\section{Modeling Methodology}

Since the mean $\Delta$$n = 0$ rate coefficients for CNO show a obvious dependence on the residual charge of the ion before recombination (Ali et al. 1991), we would expect a 
strong charge dependence among the 3rd and 4th row M-shell ions.  In Figure 1, we show the
rates for the L-Shell Fe and Ar ions from Gu (2003) for an electron temperature of 4 eV ($\sim$ 5 x 10$^{4}$K), similar to what we would expect in
photoionized plasmas in which the iron abundance was peaked at the M-shell. Since they span the same range in residual charge, 
one might expect that the M-shell Fe rates would be similar to the L-shell Ar rates. Hence, we estimated the rate coefficients for iron by 
generating 2$^{\rm nd}$-order polynomial fits to the L-shells rates of Fe and Ar and the mean rates for the CNO sequence (Ali et al. 1991), with the functional form:
\begin{equation}
R = 1.0 \times 10^{-10}~ 10^{C1 + C2xZ +C3xZ^{2}} {\rm cm}^{3}~ {\rm s}^{-1}
\end{equation}
where $Z$ is the residual nuclear charge of the ion before recombination. The individual fits are shown in Figure 1, and the 
values of the fit coefficients are listed in Table 1. 
Since Gu's calculations show a different dependence on $Z$ for different elements within the L-shell, we opted not to use this fit for the other 3rd and 4th row elements.
(in some cases, the behavior within the shell cannot be fit with a second order polynomial).
However, for the sake of completeness, we did estimate $\Delta$$n = 0$ rates 
for these ions using a linear fit to set of data including the Fe and Ar L-shell rates and the mean 
rates for the CNO sequence (Ali et al. 1991), as shown by the dotted line in Figure 1, with the functional form:
\begin{equation}
R = 3.0 \times 10^{-12}~ {\rm}  10^{0.1Z}~  {\rm cm}^{3}~ {\rm s}^{-1}.
\end{equation}
Since $\Delta$$n = 0$ DR are not fast for recombination from closed shells over the range of
temperatures we consider, we did not include estimated
DR for those ions. Note that we did not include temperature dependence for our estimated rate coefficients.

To quantify the effects of these missing DR rates we performed a series of photoionization
calculations using the best existing data, but with and without these estimates of the $\Delta$$n = 0$ DR.
We did not introduce new high-temperature DR rates, using, instead, the fit to the rates for Fe 
given in Arnaud \& Raymond (1992).
For the models, we assumed a slab geometry, solar abundances (e.g. Grevesse \& Anders 1989), a number density n$_{H}$ $=$ 10$^{8}$ cm$^{-3}$
and total hydrogen column density of N$_{H}$ $=$ 1 x 20$^{22}$ cm$^{-2}$, typical of X-ray absorbers; note that the predicted ionization structure
is insensitive to n$_{H}$, unless the densities exceed 10$^{10}$ cm$^{-3}$. For the spectral energy distribution of the incident continuum
radiation, we assumed a power law of the form
F $\propto$ $\nu$$^{-\alpha}$, where $\alpha$ $=$ 1 for $h\nu$ $<$ 13.6 eV, 1.4 for 13.6 eV $\leq$ $h\nu$ $\leq$ 600 eV, and 0.8 for
$h\nu$ $>$ 600 eV. We generated the two sets of models over a range in ionization parameter (number of ionizing photons per nucleon at the illuminated
face of the slab) log U = $-$1.0 to 1.0, in increments of 0.05 dex. 
Although the mean electron temperatures range from 1.8 x 10$^{4}$K to 4.5 x 10$^{5}$ K, the M-shell Fe ions peak between log U $=$ $-$0.5 and 0.5, or 
3 x 10$^{4}$K $\lesssim$  T $\lesssim$ 9.0 x 10$^{4}$ K, justifying the energy we used in deriving the rate coefficients. Over this temperature range, the DR 
rates may vary by factors of $\lesssim$ a few, which is within the uncertainty quoted above.
Note that, while T increased slightly when the estimated DR rates are included, the ionic columns
for 1st and 2nd row elements remained essentially unchanged. 

The predicted ionic column densities for Fe~VIII through Fe~XVI are shown in Figures 2 through 4. Since 
intrinsic X-ray absorption is typically characterized by strong oxygen bound-free edges and absorption lines, we also include the O~VI, O~VII,
and O~VIII columns for comparison. As shown in Figure 2, the ionization parameters at which the column densities of Fe~VIII, Fe~IX, and Fe~X 
peak have all shifted to higher values of U. 
In the case of Fe~X, U has shifted by a factor of $\sim$ 1.5. More importantly, the range in ionization parameter over which a significant fraction of the
total Fe exists as Fe~X (i.e., 10\%, corresponding to a column density $\gtrsim$ 10$^{16.6}$ cm$^{-2}$) has changed from $-0.8 \leq 
{\rm log U} \leq-0.3$ to $-0.8 \leq {\rm log U} \leq 0.1$. This suggests that, due to $\Delta$n $=$ 0 DR, relatively low ionization states of Fe can 
contribute significantly to opacity of more highly ionized plasmas than would be predicted using the currently available atomic data. Although the 
peak shifts are not significant for Fe~VIII and Fe~IX, 
the curves show substantially more of these ions at higher values of U. Note that, since Fe~IX has a closed shell (Ar-like), there is no
low temperature $\Delta$$n = 0$ process for recombination to Fe~VIII, which is produced primarily by radiative recombination. The increase in Fe~VIII at 
higher U is due to the larger fraction of Fe~IX, which is produced by DR from Fe~X. The structure in the high ionization parameter tails of
Fe~IX and Fe~X is the result of high temperature DR, although this effect is swamped for Fe~X in the models in which $\Delta$n $=$ 0 DR included.
Fe~XI through Fe X~III are shown in Figure 3, and for each ion, the peaks shift to higher ionization
by factors of $>$ 2, reflecting the large $\Delta$n $=$ 0 rates. Fe~XIV through Fe~XVI are shown in Figure 4, 
and the inclusion of $\Delta$n $=$ 0 DR in this case {\it decreases} the fractional abundances of these ions over much of the
adopted range in $U$, due to the large recombination rates. Nevertheless, $\Delta$n $=$ 0 models predict larger columns of these ions for 
Log U $\gtrsim$ 0.7, as a result of the faster recombination of the L-shell ions.
  
In order to illustrate the effects of uncertainties in the estimated DR rates, we introduced a ``noise factor'' by using a random
number generator with a Gaussian distribution, with a standard deviation of 0.3 dex, suggested by the largest error in the polynomial fits (see Figure 1). 
Furthermore, the DR rates computed by Gu (2003) for L-shell Ar vary by $\sim$ $\pm$ 0.3 dex over the range in
electron temperatures predicted by our models.
Each time Cloudy was run with this option, the recombination coefficient was multiplied by the random number. The scale factor was kept
constant for a particular calculation, but different random numbers were used for different calculations. We generated a set of 20 models at each 
value of U, and determined the standard deviation of predicted column densities. In Figure 5 we show the scatter for Fe~X at three values of U.
Clearly, uncertainties in the rate coefficients have an measurable effect on the predicted ionic columns. Nevertheless, the trend to peak abundance
at higher ionization parameter is still obvious, as demonstrated by the overplot of the Fe~X columns from the models without the 
estimated DR.
  
\section{Conclusions}

The results of this simple test indicate that the model predictions of the Fe ionization balance depend critically on accurate $\Delta$$n = 0$ 
DR rate coefficients. Photoionization models of X-ray absorbers that match spectral features from abundant 2nd row elements such as
C, N, O, and Ne will overpredict the average Fe ionization state if this process is not included. This can be easily demonstrated using Figures 1 -- 3, by selecting an ionization
parameter to give the proper ratio of the O~VI, O~VII, and O~VIII column densities and reading off the relative Fe ionic columns for
the models with and without the estimated DR.  The apparent discrepancy in the ionic states detected in the Fe M-shell
UTA in NGC 3783 (Netzer et al. 2003), compared to those predicted by the absorber models, is likely due to the absence of $\Delta$$n = 0$ rates.
To illustrate this point, we regenerated the so-called low ionization model used by Netzer et al. to fit the soft X-ray absorption
and inner shell Si and S lines detected in a 900 ks merged {\it Chandra}/High Energy Transmission Grating spectrum of NGC 3783 (see, also, Kaspi et al. 2002).
Following Netzer et al., we assumed the SED described in Kaspi et al. (2001) and set log U $=$ $-$0.26 and N$_{H}$ $=$ 10$^{21.9}$ cm$^{-2}$ (note that our  
conversion from the X-ray ionization parameter used by Netzer et al. to the form of U used in CLOUDY resulted in a model prediction of slightly higher
relative ionization of Si and S than they predicted, which may be due to slight differences in the model SED; in order to reproduce the same columns of
Si and S we adjusted the value of U downward by 10\%). The predicted ionic columns for Fe~VIII through Fe~XII, without $\Delta$n$=$0 DR, are given in Table 2, 
and the combined Fe~XI and Fe~XII column is $\sim$ 3.5 greater than the combination of Fe~VIII and Fe~IX, while the peak ionization state
is Fe~XI. Since the strongest UTA features for Fe~XI and Fe~XII lie at 16.16\AA~ and 15.98\AA~, respectively (Behar et al. 2001), the UTA would peak near 16.1\AA~, as 
described by Netzer et al.. Also shown are the predictions for the same model with $\Delta$n$=$0 DR included. In this case, the peak ionization state is Fe~X,
and the total Fe~VIII and Fe~IX column are approximately equal to those of Fe~XI and Fe~XII combined. Since Fe~VIII and FeIX produce their strongest 
features at 16.89\AA~ and 16.51\AA, respectively (Behar et al.), and the profile would be dominated by Fe~X (16.34 \AA), the predicted UTA will peak near
16.4\AA~, as observed. It should also be noted that including the new L-shell rates for 3rd row elements such as S, Si, and Mg will alter the predicted
ionization balance for those elements as well. As noted by Behar \& Netzer (2002), inner shell absorption lines from these ions can be used
as effective probes of high-column absorbers. However care must be taken in using photoionization models to determine ionization states and total
column densities of the absorbers using these lines as constraints, at least until a full set of accurate L-shell DR rates have become available.

Seyfert galaxies are also strong soft X-ray emission-line sources (e.g., Sako et al. 2000; Kinkhabwala et al. 2002; Turner et al. 2003). It will be
important to understand the connection between this component and forbidden ``coronal'' emission lines such as [Fe~X] $\lambda$6374, 
[Fe~XI] $\lambda$7892, [Fe XIV] $\lambda$5303 and [S~XII] $\lambda$7611 (e.g. Kraemer \& Crenshaw 2000), especially
since the optical/UV spectral images obtained with the {\it Hubble Space Telescope}/Space Telescope Imaging Spectrograph provide
tighter constraints on the spatial extent and kinematic properties of the emission-line gas. Accurate DR rates are crucial for modeling this component and
exploring the link between the X-ray and optical emission lines.

While the models generated for this Letter clearly illustrate the magnitude of the problem, we are left with the question on how best to proceed
in the absence of a full set of $\Delta$$n = 0$ DR rate coefficients.  
Our Monte Carlo 
calculations (see Figure 5) provide an estimate of the error in the predicted column densities due to the uncertainties in the
rate coefficients. In fact, our results suggest that comparisons between calculated and observed column densities be given error bars
$\sim$ 0.3 dex, assuming our rates are included in the calculations, due to the uncertainties in the rates and the unmodeled temperature 
dependence (as suggested by the variations in the 
published L-shell rates over the range in temperature predicted by our models, e.g.,  1.8 x 10$^{4}$K $\leq$ T $\leq$ 4.5 x 10$^{5}$ K). 
Clearly, using rates 
estimated from other ions is far better than assuming the process does not occur, which is what is done 
when it is neglected. Hence, we suggest that it is best to adopt these estimated rate coefficients with this stated error.
Ultimately, new atomic data, in general, and DR rates, in particular, are needed to reliably compare the ionization of third
and fourth row elements.

\acknowledgments

 S.B.K. and J.R.G. acknowledge support from NASA grant NAG5-13109. S.B.K. thanks Hagai Netzer for
 useful discussions concerning the Fe UTA lines. We thank an anonymous referee for useful comments.

\clearpage

\figcaption[f1.eps]{The $\Delta$n $=$ 0 DR rate coefficients for L-shell Fe ions (crosses), L-shell Ar ions (stars), and the mean
rates for the 3rd and 4th ionization states of the CNO ions (diamonds) are shown as a function of the residual charge
of the ion before recombination.
The L-shell rates are those from Gu (2003), for a temperature of 4.0 eV. The dashed-lines are 2$^{nd}$-order polynomial
fits for the three shells (N-, M-, and L-), which we used as the functional form for our estimated Fe rate coefficients.
The dotted line is a linear fit to the rates, which we used as the functional form for the other 3rd and 4th row elements.}

\figcaption[f2.eps]{Predicted ionic column densities for Fe~VIII (dotted line), Fe~IX (dashed line)
and Fe~X (dash-dotted line) are shown as a function of ionization parameter (U). The blue-lines and red-lines are the 
predictions with and without the
estimated $\Delta$n $=$ 0 rates, respectively. The solid black lines are the predicted column
densities for O~VI, O~VII, and O~VIII, as indicated. For both sets of models, we assumed solar abundances and a 
hydrogen column density N$_{H}$ $=$ 10$^{22}$ cm$^{-2}$.}

\figcaption[f3.eps]{Predicted ionic column densities for Fe~XI (dotted line), Fe~XII (dashed line)
and Fe~XIII (dash-dotted line) are shown as function of U. All other symbols
are as defined for Figure 3.}

\figcaption[f4.eps]{Predicted ionic column densities for Fe~XIV (dotted line), Fe~XV (dashed line)
and Fe~XVI (dash-dotted line) are shown as function of U. All other symbols
are as defined for Figure 3.}

\figcaption[f5.eps]{The predicted Fe~X column density is shown as a function of U. The solid-line and dashed-lines are the 
predictions with and without the
estimated $\Delta$n $=$ 0 rates, respectively. The error bars are included to show the effect of 
uncertainties in the DR rates, which we modeled by introducing a random ``noise
factor'' (see text).}

\clearpage

\vskip3.0in
\begin{figure}
\psfig{file=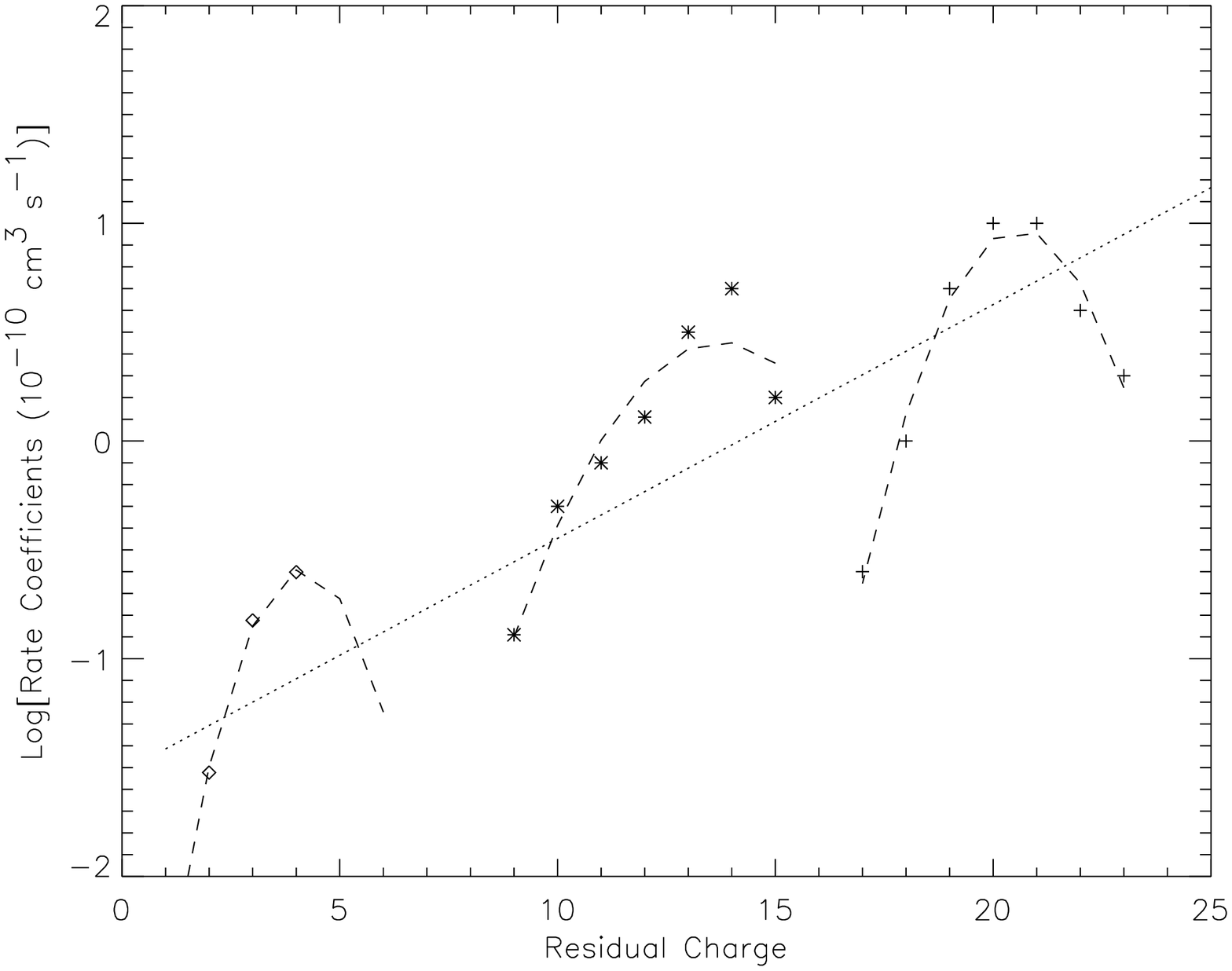,width=5.5in,angle=90}
\end{figure}

\clearpage
\vskip3.0in
\begin{figure}
\psfig{file=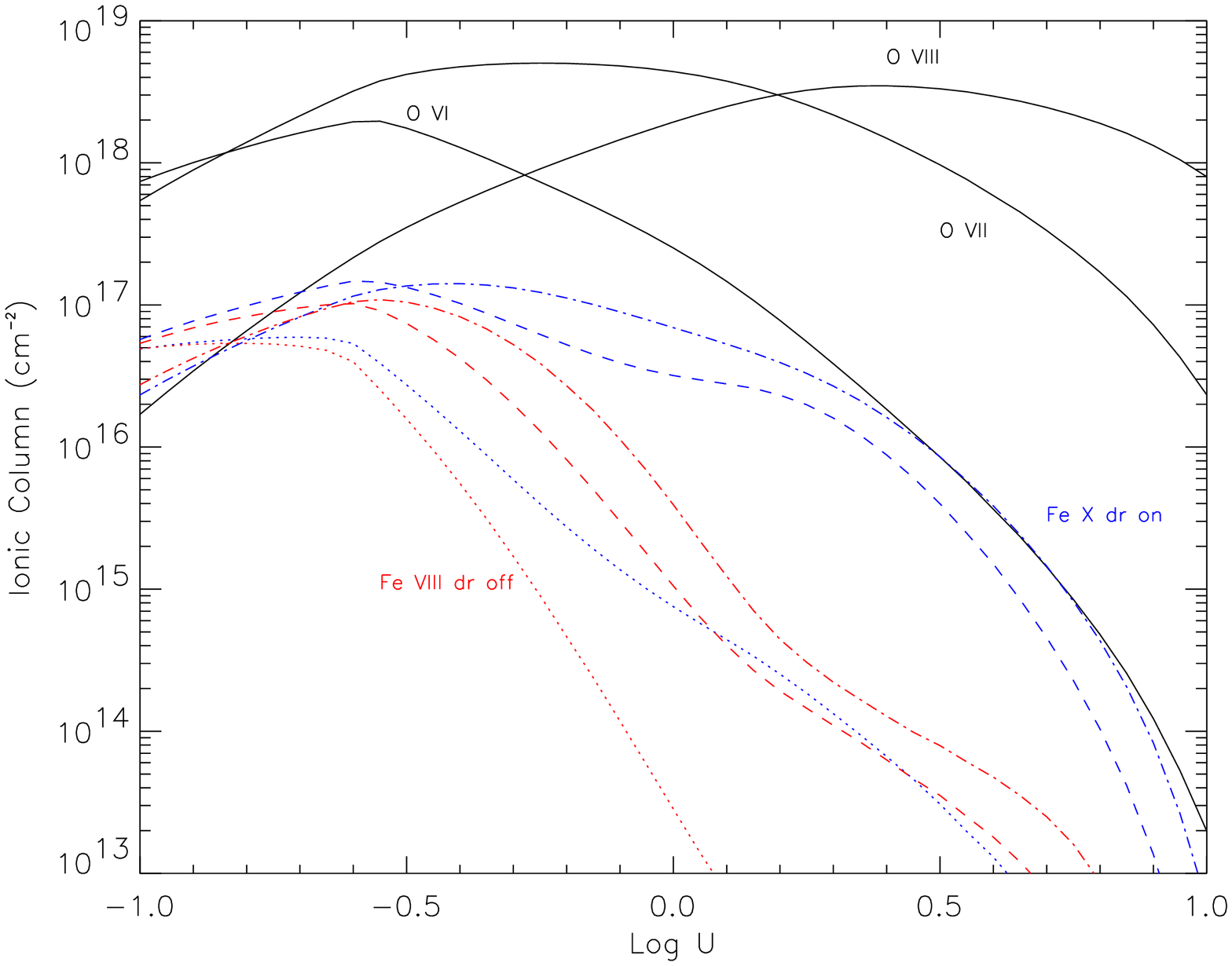,width=5.5in,angle=90}
\end{figure}

\clearpage
\vskip3.0in
\begin{figure}
\psfig{file=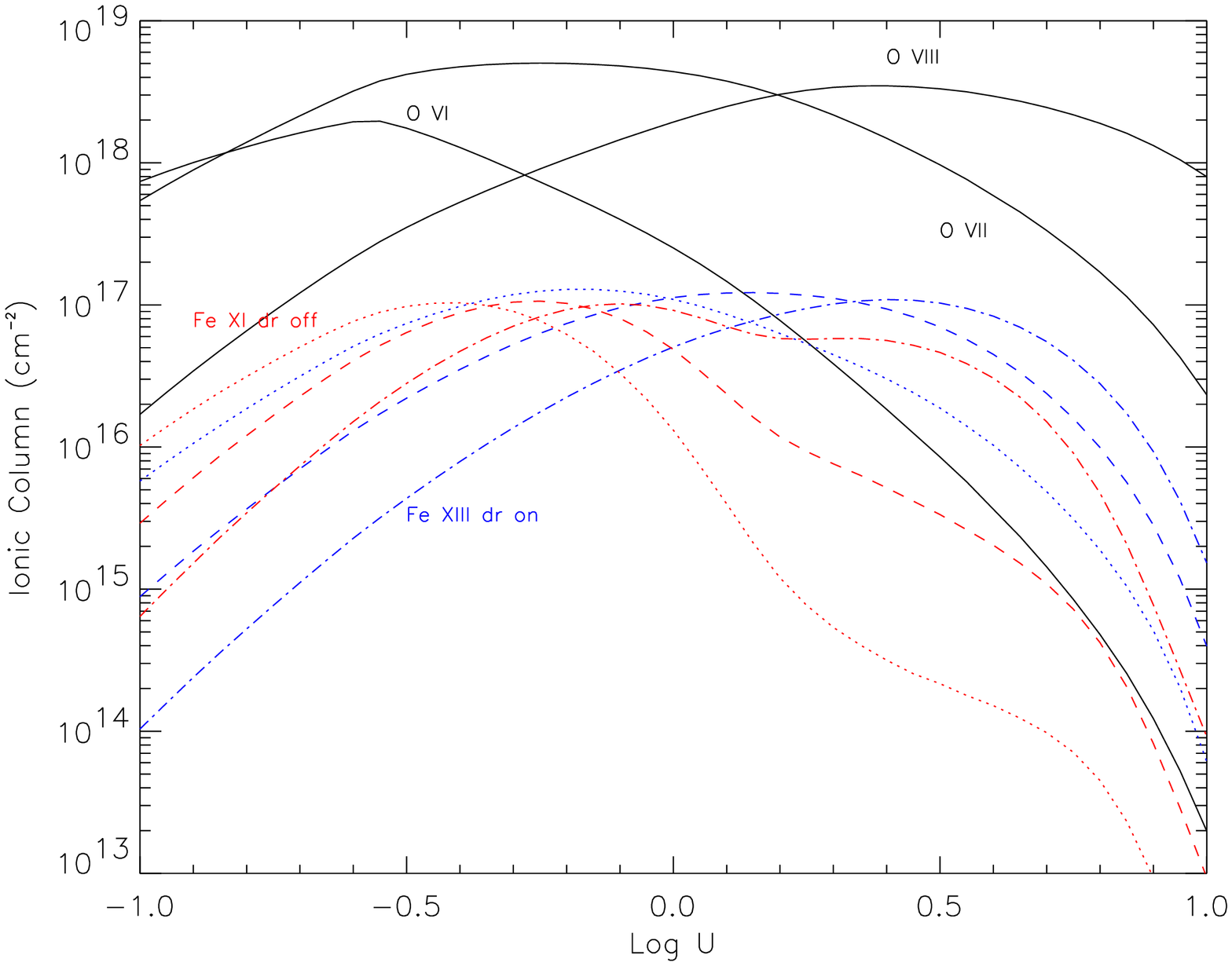,width=5.5in,angle=90}
\end{figure}

\clearpage
\vskip3.0in
\begin{figure}
\psfig{file=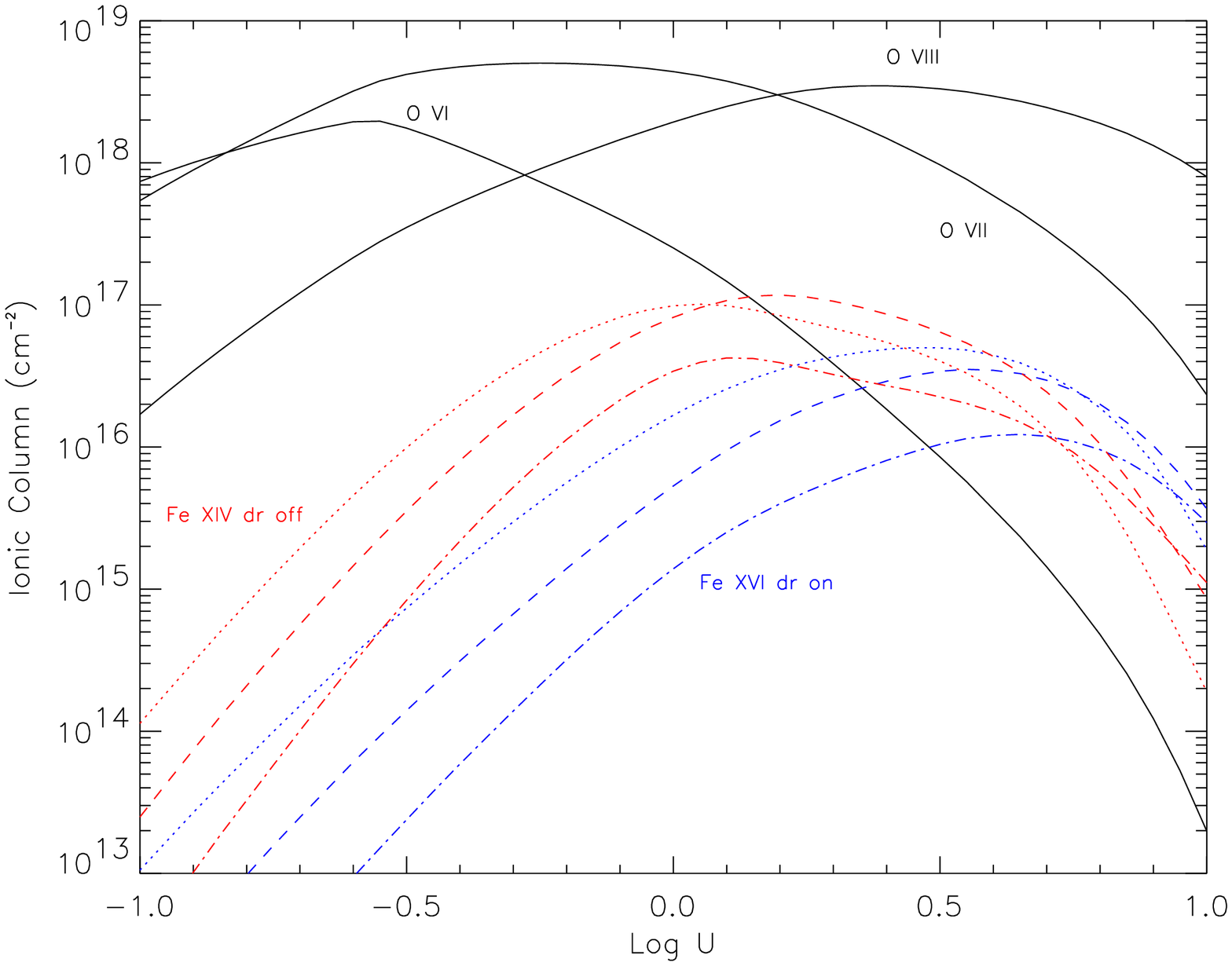,width=5.5in,angle=90}
\end{figure}

\clearpage
\vskip3.0in
\begin{figure}
\psfig{file=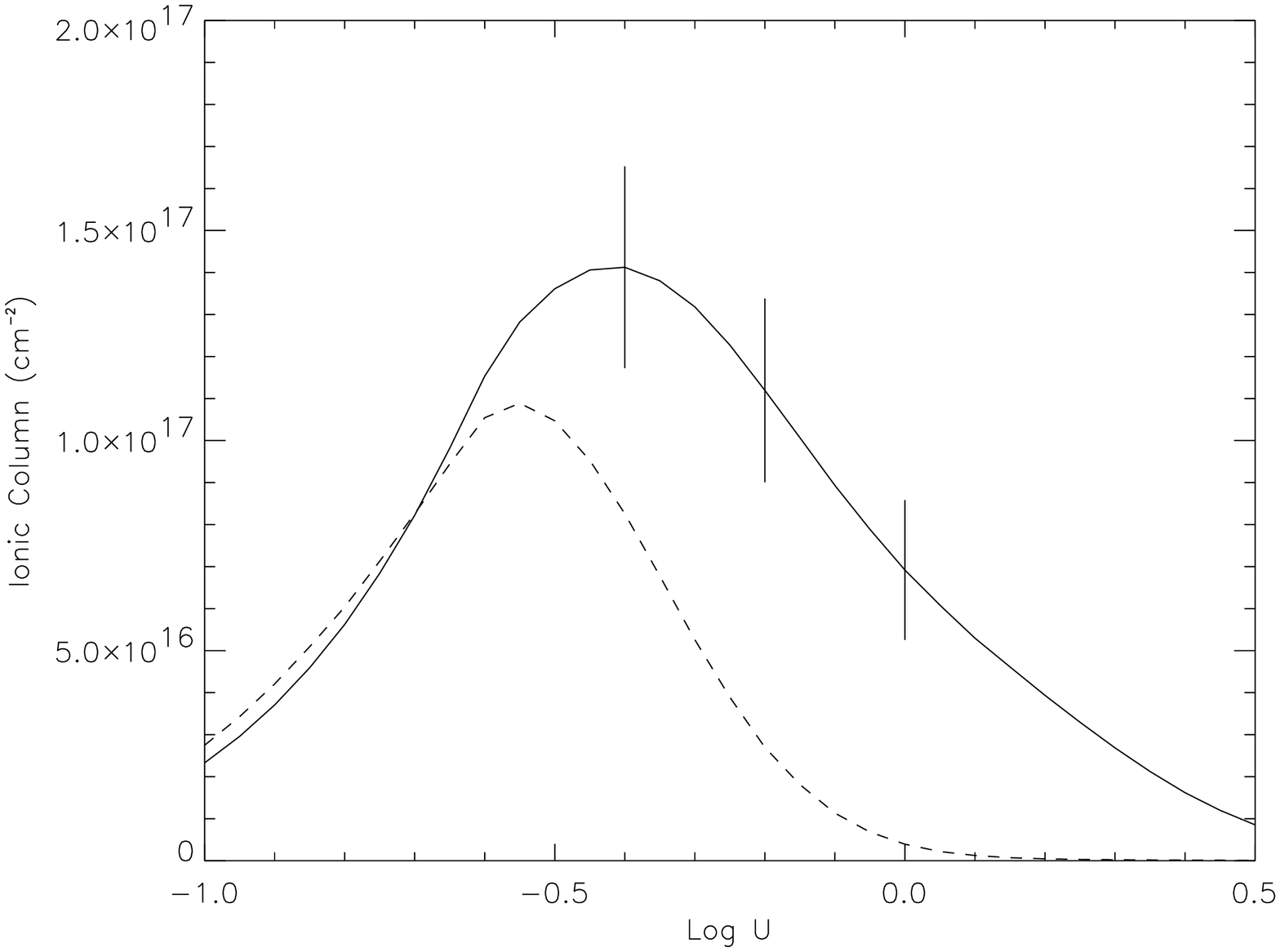,width=5.5in,angle=90}
\end{figure}

\clearpage

\begin{deluxetable}{lccc}
\tablecolumns{4}
\footnotesize
\tablecaption{Polynomial Fit Coefficients for the Estimated Fe DR rates}
\tablewidth{0pt}
\tablehead{
\colhead{Z} & \colhead{C1}  & \colhead{C2} & \colhead{C3}}
\startdata
17 -- 23 & $-$52.5073 & 5.19385 & $-$0.126099 \\
9 -- 15 & $-$10.9679 & 1.66397 & $-$0.060596 \\
$<$ 8 & $-$3.95599 & 1.61884 & $-$0.194500 \\
\\
\enddata
\end{deluxetable}
 
\clearpage

\begin{deluxetable}{lcc}
\tablecolumns{3}
\footnotesize
\tablecaption{Predicted Fe M-Shell Column Densities (cm$^{-2}$) for NGC 3783$^{a}$}
\tablewidth{0pt}
\tablehead{
\colhead{Ion} & \colhead{log(N$_{ion}$), no $\Delta$n $ =0$ DR} & 
\colhead{log(N$_{ion}$), with $\Delta$ N$=0$ DR}} 
\startdata
Fe~XIII & 16.62 & 15.90 \\
Fe~XII & 16.83 & 16.48\\
Fe~XI & 16.88 & 16.88\\
Fe~X & 16.78 & 17.02 \\
Fe~IX & 16.52 & 16.90 \\
Fe~VIII & 15.81 & 16.18\\
\\
\tablenotetext{a}{Predictions are for the low-ionization component of the X-ray absorber
described in Netzer et al. (2003), assuming: log(U) $=$ -0.26, log(N$_{H}$) $=$ 21.90.}
\enddata
\end{deluxetable}

\end{document}